# How Diverse Users and Activities Trigger Connective Action via Social Media: Lessons from the Twitter Hashtag Campaign #ILookLikeAnEngineer

Aditya Johri
George Mason University
johri@gmu.edu

Habib Karbasian
George Mason University
hkarbasi@gmu.edu

Aqdas Malik
George Mason University
amalik21@gmu.edu

Rajat Handa
George Mason University
rhanda@gmu.edu

Hemant Purohit
George Mason University
hpurohit@gmu.edu

## Abstract

*We present a study that examines how a social media activism campaign aimed at improving gender diversity within engineering gained and maintained momentum in its early period. We examined over 50,000 Tweets posted over the first ~75 days of the #ILookLikeAnEngineer campaign and found that diverse participation – of types of users – increased activity at crucial moments. We categorize these triggers into four types: 1) Event-Driven: Alignment of the campaign with offline events related to the issue (Diversity SFO, Disrupt, etc.); 2) Media-Driven: News coverage of the events in the media (TechCrunch, CNN, BBC, etc.); 3) Industry-Driven: Web participation in the campaign by large organizations (Microsoft, Tesla, GE, Cisco, etc.); and 4) Personality-Driven: Alignment of the events with popular and/or known personalities (e.g. Isis Anchalee; Michelle Sun; Ada Lovelace.). This study illustrates how one mechanism – triggering – supports connective action in social media campaign.*

## 1. Introduction

In August 2015 a hashtag on Twitter, #ILookLikeAnEngineer, started trending highly and attracted a lot of media coverage and participation by users. The hashtag was born out of one woman's frustration with how people reacted online to an offline recruitment advertisement by her company OneLogin™ that featured her on billboards on public transport areas. On the Web, many users, mostly male, reacted to the billboard by saying she cannot really be an engineer or that the company used the picture of a female engineer solely to attract male applicants. The engineer featured in the billboard, Isis Anchalee, wrote a blog post about her experience that went viral and called for supporters to coalesce around her efforts to broaden the depiction of what an engineer looks like, hence the hashtag #ILookLikeAnEngineer.

Diversity of workforce is critical for designing a broad range of products that target different user based. It is also increasingly evident that given the role of engineering, computing, and technology, in everyday life a more diverse workforce is needed by companies to remain competitive in the future by recruiting the most capable workforce [26, 27]. Yet, efforts, including policy making [30, 31, 32], to encourage and support women's participation into STEM workforce are consistently undercut by the inherent and implicit bias against women, as the reaction to a billboard featuring a women engineer illustrates. The representation of women in U.S. engineering education and workforce has stayed remarkably low at less than 25% [28, 29]. In Silicon Valley, where the hashtag campaign we examine started, the gap is even wider, especially in executive suites and corporate boards.

In this paper we present a study that looks closely at the #ILookLikeAnEngineer campaign to better understand the multifarious manner in which the Web facilitates activism. We studied the initial period of the campaign to better understand what created the initial momentum and how the campaign was sustained after the first two weeks when the campaign went viral. Although it is hard to gauge the longer term impact of a campaign whose purpose is to create awareness, there are some indicators of its success. The hashtag is still in use regularly and in the first year saw over 250,000 tweets and events across the globe. Google featured it in its assessment of the year's highest searches and recognized that the search phrase "girls who code" increased more than 45% in 2015 due to the campaign.

We draw on recent theoretical work within IS as related to two aspects of social media use and research: 1) affordances and 2) connective action. Social technologies have unique affordances to support diverse participation in social movements and allow users to engage with a social movement in ways that are different than participation in earlier social movements. In particular, users can now connect loosely – through signaling their support via a Tweet, for instance – allowing for different forms of organizing or connective action. The findings from this study contribute to this work by delineating an affordance – triggering – that provides momentum to connective action during the initial phase by allowing for contributions from a broad range of actors, across a diverse range of activities, both online and offline.

## 2. Literature review

### 2.1. Social movements and women's rights

In his review of what constitutes social movements Diani [5] proposes that most definitions share three criteria: (1) there is a network of informal interactions among a diverse group of individuals, groups and/or organizations; (2) this network is engaged in some form of a political or cultural conflict; and (3) there is an accepted sense of shared collective identity among the network. Tilly [17], adding a dynamic element to this, argues that social movements, in addition to being a network of actors, are a series of contentious performances, displays and campaigns such as processions, vigils, rallies, demonstrations, petition drives, statements to media, and so on. James and van Seters [8], combine these elements together and argue that there has to be formation of a collective identity, development of a shared normative orientation, sharing of a concern for change of the status quo, and occurrence of moments of practical action that are at least subjectively connected together across time addressing this concern for change.

The case we present here has a lineage within the realm of prior work on social movements and women's rights. In the U.S., starting with the suffrage movement, a history can be traced for women's right that moves through the 60s and 70s leading to the 90s "new social movements" and post-feminism. Within the 'social movements' literature, women's right and feminism have garnered significant research and many of the theoretical developments have relied on studying these movements to further our understanding [12, 16]. In spite a long trajectory of social movements aimed at improving women's rights, there has been limited impact on women's role in broadening participation in the workforce, especially in Science, Technology, Engineering, and Mathematics (STEM) fields. The #ILookLikeAnEngineer therefore commonly shares issues of equality, diversity, and justice with prior efforts and displays many elements that constitute a social movement in itself: it relies on informal interactions among supporters, the actors who participate are engaged in a common social justice cause; and there is a sense of a commonly shared identity.

### 2.2. Digital activism and hashtag activism

The infusion of digital technologies has changed the landscape of activism and digital activism – the use of various forms of digital technologies (e.g. blogs, videos, podcasts, email, and social media) for activism purposes – has gained significant momentum in recent years. Digital technologies enable activists in disseminating information quickly to a larger audience through a multitude of channels for a social, political, economic, or environmental change/cause [9,13]. Activists use networked technologies not only for creating and sharing information but for forming public opinion, planning and calling for action, protect activists, as well as mobilizing both online and offline resources [2,3,4,14,20]. Similar to offline activism campaign, most of the digital activism campaigns are either reactive or call for a proactive action [20]. Reactive campaigning (e.g. regime change, government accountability, civil disobedience) is often politically motivated and targeted against certain controls or the authorities imposing those controls, meanwhile the proactive campaigns (e.g. disaster resilience, human rights, anti-bullying, and racial equality) promote or highlight a cause with an aim for achieving an certain objective [2,4,13,20]. Digital activism has not only been used in developed countries but increasingly in the developing ones for resistance acts and organizing public protests [2,3,6,7,10]. Contrary to offline activism campaigns, digital activism campaigns are not bound by time or place. Online users can participate and engage without these restrictions to support the cause or get involved in registering their protest [13].

Within the realm of digital activism, Internet-focused repertoires [15] have become increasingly common and the rise of social media has provided a new venue for networks to come and act together [19]. One popular repertoire that has emerged within digital activism is hashtag activism which involves large number of postings by using a common hashtagged word on social media platforms [21]. Hashtags were initially introduced by Twitter to classify tweets into common theme or topics to facilitate easy search of specific messages and have been adopted by many other social media platforms including Facebook and Instagram [18]. Hashtags have evolved over time and they are not only used for categorizing content but tailored and crafted by users for various purposes such as events, branding, breaking news, and supporting causes. In recent years, social media in general, and Twitter in particular has seen a surge of usage and is commonly associated with furthering a number of social and political issues [2,4,6,7,13, 24], e.g. hashtags such as #BlackLivesMatter, #YesAllWomen, #OccupyEveryWhere, and #BringBackOurGirls are regarded as some of influential cases of hashtag activism [21]. In the rest of the paper when referring to the #ILookLikeAnEngineer case we use social movement, online campaign or activism, and hashtag activism interchangeably – as has been the case when it has been referred to by its supporters and the media – but we are always referring to the broader offline and online campaign and not just to its online instance. In some ways, the campaign morphed into a larger movement over time.

## 2.3. Social media affordances and connective action

The study of social media has found significant traction within the communication and information systems fields in recent years. From a theoretical perspective, the two lenses relevant for this study are 1) the affordances perspective [36, 37], and 2) the connective action perspective [14, 33].

The affordance lens is derived from the work of James Gibson [39], an ecological psychologist, and has formed the foundation for the field of human-computer interaction (HCI) and at a larger level for the design of any object [40]. Affordance refers to the potential for action that a technology or object provides to users e.g. a door handle. Although a user has choice in how to use any technology, the design of that technology is already imbibed with features that afford or constrain certain actions (e.g. whether a door can be pushed or pulled). Affordance lies at the interaction of technology and the action it allows users; when individuals perceive that designed and perceptible features allow them to perform specific actions, the technology can be said to provide an "affordance" [37].

The affordance lens has largely been used to examine the use of social media within organizations and according to [37] it has specifically allowed the examining of the diffusion of social media in organizations, the use of social media, and the organizing processes occurring around and through social media. Within organizations, "the affordance lens enables us to surface issues surrounding social media that mark them as distinct technologies from the many other computerized communication technologies that have been used in organizations for the past half century, while simultaneously showing how they become constitutive features of organizational action" [37].

For any form of social activism, especially social movements, the ability to organize and affordance for people to participate is a necessity. Prior work on social movements has examined these issues to argue that collectivity – collective action and collective identity – are central attributes of successful social movements. Recent work that has looked at how people organize in the digital age argues that rather than a strong collective reaction to an issue or identity, it is the ability to connect, even loosely, that drives online activism,

"The emerging alternative model that we call the logic of connective action applies increasingly to life in late modern societies in which formal organizations are losing their grip on individuals, and group ties are being replaced by large-scale, fluid social networks [33]." Furthemore, they argue that, "These networks can operate importantly through the organizational processes of social media, and their logic does not require strong organizational control or the symbolic construction of a united 'we'. The logic of connective action, we suggest, entails a dynamic of its own and thus deserves analysis on its own analytical terms [33, pg. 748]."

In a recent article, Vaast et al. [36] combine these two perspectives to argue that, "Affordances of social media for connective action may thus be actualized by emerging and fluid groups of actors who are involved in different ways and to different degrees in connective action (p.5)" The authors further suggest that different users might take on different roles within the activity "complementary, interdependent roles that make up the connective action." These relationships, according to them, have not been investigated empirically and there is a need to better understand the affordances of social media that enable connective action [36, pg. 5-6]. Using a case study of the Gulf Oil Spill and activity generated around that event on Twitter, the authors argue for the concept of "connective affordances" which they define as "collective level affordances actualized by actors in team interdependent roles." They found that different users take on one of three roles – advocates, supporters, or amplifiers – and their interdependence plays a part in the overall activity. According to the authors, the concept of connective affordances extends research on affordances as a relational concept. It does so by considering not only the relationships between technology and users but also the interdependence among users and the effects of this interdependence on what users can do with the technology. They also argue that the patterns of feature use for emerging groups of actors are intricately and mutually related to each other.

In this study we further examine the relationship between affordances and connective action. In our case the interdependence among users was limited, they were bound together largely by their interest in the topic, yet their participation resulted in sustained momentum beyond the initial period. We argue that a network of unrelated events created this moment triggered both by Twitter activity but also offline activities and even those online activities that took part in forums other than Twitter.

## 3. #ILookLikeanEngineer campaign

The #ILookLikeAnEngineer Twitter hashtag was an outgrowth of an advertising campaign by the company OneLogin. In late July 2015 the company OneLogin posted billboards across public transport in the California Bay Area, especially at the BART train stations, as a recruitment campaign. The billboards depicted different engineers working in the company at that time. In this series of billboards, the engineers all wore black t-shirts and were photographed from their head till about their waist. The billboards also exhibited the engineer's name, job description, and a very short quote alongside the photo. One of the billboards depicted a woman engineer named Isis Wenger (now Isis Anchalee and that is how we refer to her in the paper) and her photo attracted a lot of attention on the Web (Figure 1).

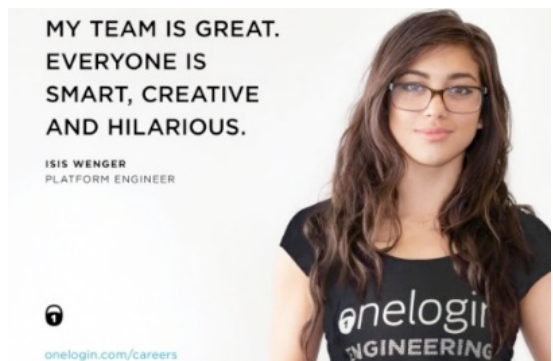

**Figure 1: Billboard campaign featuring Isis Anchalee**

Her image led to discussions online about the veracity of the campaign as some people found it unlikely that she was really an engineer. Online comments stated that she was "too attractive" to be a "real engineer," among other demeaning comments. These online discussions, according to Anchalee, prompted her to write a post on Medium, on August 1, 2015, responding to the stereotypical reaction her image in the ad had generated. She stated in the post that, "At the end of the day, this is just an ad campaign and it is targeted at engineers. This is not intended to be marketed towards any specific gender — segregated thoughts like that continue to perpetuate sexist thought-patterns in this industry."

On August 3, 2015, in addition to a Tweet she updated her initial blog post to add a call for action and an image with a Twitter hashtag: Do you feel passionately about helping spread awareness and increase tech diversity? Do you not fit the "cookie-cutter mold" of what people believe engineers "should look like?" If you answered yes to any of these questions I invite you to help spread the word and help us redefine "what an engineer should look like" #iLookLikeAnEngineer" (Figure 2).

She wanted to show that she was not the only female engineer and also take some of the attention away from her and towards the issue of diversity in engineering. "When this is all in the past, I would like to have created a larger impact on the community than simply generating a Twitter presence. Spreading awareness is the first step, but I want to help facilitate concise plans of action so we can create a genuine change. If you have any personal input and advice that you think could help make a difference please feel free to share it with me. You can help support us by checking out [link to a new billboard campaign followed by link to an event]".

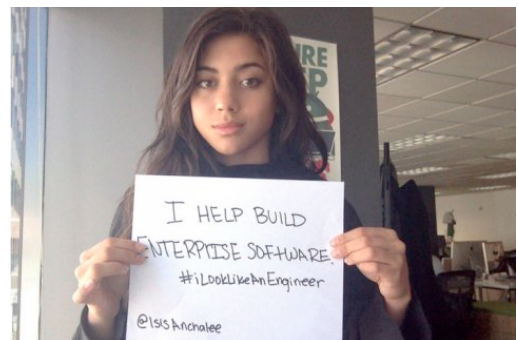

**Figure 2: Start of Twitter campaign**

Her Medium post as well the hashtag received large support and the Twitter tag in particular took off as a way for others to share their images and their ideas on the issue. The hashtag soon saw significant media coverage not just in the US but also across the Atlantic, particularly in the UK. The movement not only had an online presence but a large event was organized in September 2015 in the Bay Area supported by a group. The hashtag movement was started as a campaign to raise funds to be able to post billboards across the Bay Area with the hashtag and photos; in effect, to take the online movement into the offline world through a series of billboards. Overall, the use of the Twitter hashtag, combined with media coverage and the decision to post billboards across Silicon Valley turned the effort into a larger scale activism effort, in essence, a movement for recognizing women's participation in engineering.

## 4. Research study

In our study, we aim to answer the following research questions in this analysis of how the Web facilitates the sustainability of online activism in the early stages:

a. What is the participation pattern of diverse actors in the #ILookLikeAnEngineer campaign?

b. How does the diverse actor participation vary by time and location as the campaign progresses?

c. What are the triggers that provide momentum and sustain the campaign in the early stages?

### 4.1. Data description

We collected the dataset from Twitter using Streaming API, and based on three hashtags – #ILookLikeAnEngineer, #LookLikeAnEngineer and #LookLikeEngineer – as we had found instances of all of them being used in conjunction. The time frame for the data ranges from August 3rd, 2015, the day the hashtag was first used, till October 15th, 2015, which is about 2 months after the first initial surge of the campaign. The dataset consists of 19,354 original Tweets and 29,529 Retweets. For each Tweet we collected the following metadata: text, retweet count, favorite count, time of tweet; for the user: name, screen name, location, followers, following, likes, number of statuses.

### 4.2. Preliminary analysis

Table 1: Tweet/Retweet analysis

|  | Tweet | Retweet | Favorite | Following | Followers | Likes |
|---|---|---|---|---|---|---|
| Tweet | 1 |  |  |  |  |  |
| Retweet | 0.177 | 1 |  |  |  |  |
| Favorite | 0.208 | 0.881 | 1 |  |  |  |
| Following | 0.013 | 0.023 | 0.022 | 1 |  |  |
| Followers | 0.013 | 0.331 | 0.234 | 0.044 | 1 |  |
| Likes | 0.016 | 0.020 | 0.030 | 0.103 | 0.009 | 1 |
| Statuses | 0.029 | 0.003 | -0.005 | 0.077 | 0.051 | 0.099 |

In order to find the main contributors to this cause, we focused only on the metadata of original tweets (~19.5K). Our preliminary analysis indicated that the correlation between Favorite and Retweet is the strongest meaning that a Tweet is Favorited as many times as Retweeted (Table 1) indicating that some Tweets within the overall dataset have significantly more impact in terms of awareness and reach than others. In particular, Tweets and Retweets by media organizations or, in this case, those by known technology leaders, had much higher Retweet and Favorite count as compared to the average Tweet. We discuss this further in the findings section.

Overall, the participation was from 138 countries with a majority, 64%, of tweets were from the US. Statistics of the tweets per state point out that most of the tweets come from states where mostly high tech companies are located in East and the West coast, especially California where many startup companies are located. Isis Anchalee has the highest favorite count. Broadly, the top contributors can be divided into individuals, companies such as Tesla, Intel, GE, and news and media organizations such as BBC, Independent, BuzzFeed, Fortune, TechCrunch (Table 2). The most active user is the female engineer (Michelle Glauser), who was the manager for the hashtag campaign and who worked with Anchalee to start the campaing. Although the number of her tweets is many, the retweets and favorites she received were also high as her followers were high as well.

Using Degree Centrality, we calculated the in-degree and out-degree for each actor in our network and then sorted out the results to get the prominent and sociable users in the network, respectively. The analysis revealed that 76% of prominent users were organizations. PageRank centrality and Betweeness further confirmed that organizations played an important role in the trending of the movement and in connecting and bringing users together for movement. Overall, we observed a markedly strong role for organizations in the campaign, which is likely due to their reputations and workforce diversity related initiatives.

## 5. Triggers of increased activity

Our preliminary analysis of the data brought to light many interesting aspects of the campaign. It highlighted the prominent role of the industry and different technology companies in the campaign. It also brought to light the major individual users and the interactions between different user populations. The analysis also made it transparent that the community behind the movement was not necessarily a tightknit network but that at different point different actors and activities played a prominent role, including media organizations. From a geographic perspective, the movement was concentrated in expected areas – those with a higher presence of the engineering and technology companies and especially the California Bay Area, Silicon Valley, where the campaign originated.

Our preliminary analysis guided us to look for specific events, users, and media coverage to better understand the triggers of initial momentum through increased activity ("buzzmakers" as [24] refers to them) and as we proceeded with an in-depth examination of our data corpus, we were able to clearly demarcate their role in the campaign. As a first step, an overall temporal analysis was plotted of the original tweets, retweets, and favorites (see Figure 3). This plot clearly showed the high initial momentum of the campaign and then the relatively slide into lower activity. Across our dataset the averages are as follows: Tweet posts = 264 Original Tweets per day; Retweet = 1211 per day; Favorite = 1930 per day. Therefore, even though the activity is low, it is not insignificant and the expanded view of the temporal plot shows the peaks and troughs in the data. As a next step three members of the research team independently sifted through the entire dataset and coded for a list of prospective triggers following an open coding methodology where all instances that lead to a spike in participation were recorded. The three lists were then compared to create a composite list of common triggers.

Next, the triggers were categorized into a final list of four: (1) Event-Driven; (2) Media-Driven; (3) Industry-Driven; and (4) Personality-Driven. The category industry-driven was conceived to be broad and included for-profits and non-profits, including academic institutions. It was clear even from our preliminary qualitative analysis that these triggers often worked in conjunction with each other. For instance, an event garnered support not just through participation in the physical activity but also through its media coverage, especially if it was reported in media outlets that serve the technology industry. Personalities not only brought on board their supporters but also their parent organizations. We next discuss each trigger in detail.

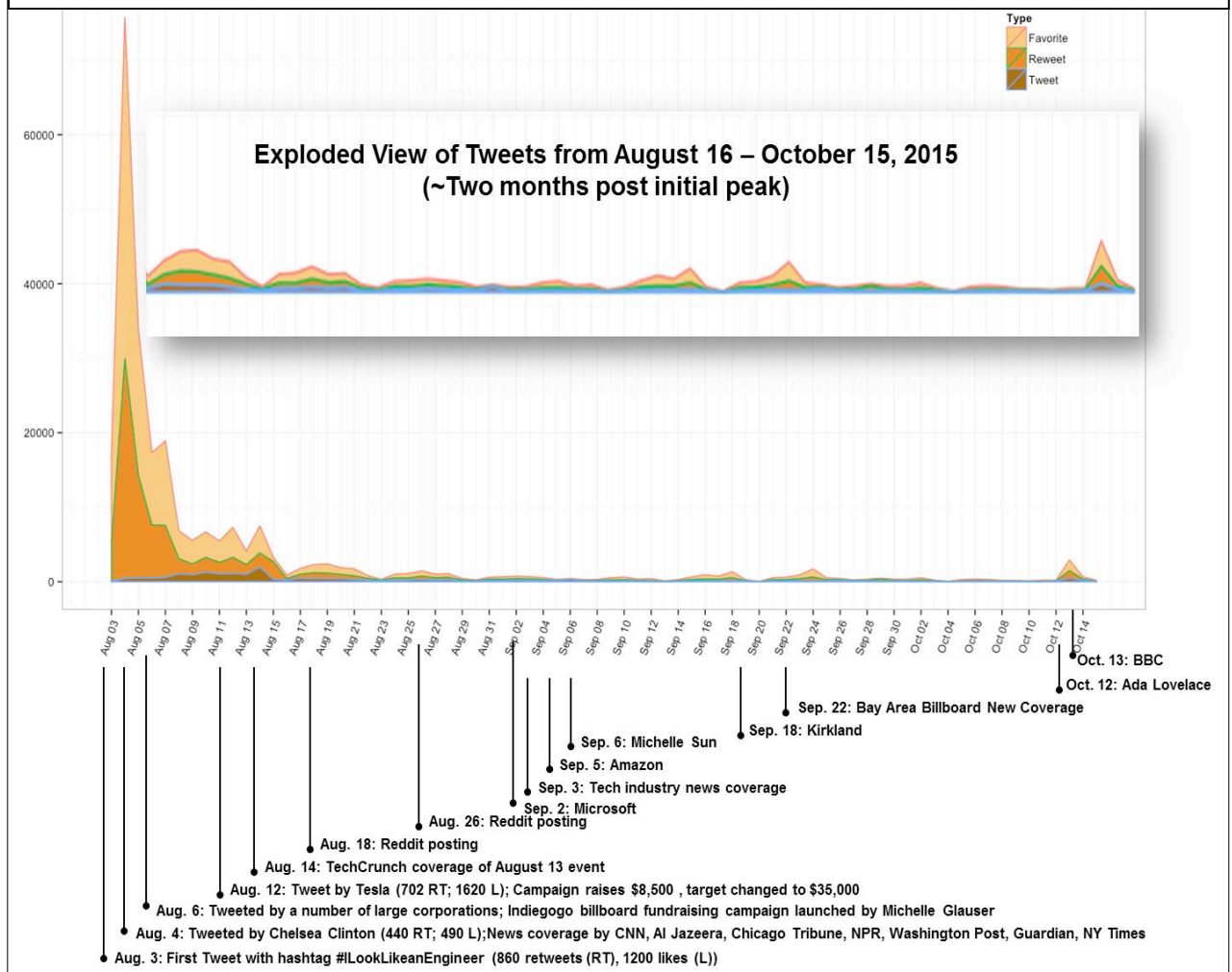

Figure 3: Temporal representation of activity triggers over the dataset with expanded view of Aug. 16-Oct. 15

## 5.1. Event-driven triggers of activity

The prominent role played by offline events in calls for action related to the campaign is evident from its inception. As early as August 6th, 2015 a message about holding an offline event in San Francisco was posted on a blog and other online channels through the hashtag:

> "This event, inspired by the negative attention received by engineer Isis Wenger and the resulting inspiring #ILookLikeAnEngineer Twitter movement, will bring together underrepresented engineers (women, PoC, LGBT, non-white-or-asian-straight-cis-males) for an evening to financially support the creation of billboards highlighting these engineers and to give us a chance for discussion."

The event that took place on August 13 brought a massive surge in participation, primarily through its coverage in numerous media outlets. This event also had an online component in the sense that participants at the event were photographed to become the face of the online campaign and through a physical billboard campaign to be place across Silicon Valley. The associated online campaign to collect funds for the billboard campaign led to more activity. A few weeks later, Anchalee was invited to be part of TechCrunch Disrupt, another physical event, and similar to the previous event this one was covered by the media and led to surge in activity. In addition to these larger and more visible events, there were also smaller local events that resulted in increased activity e.g. small group of engineers got together to promote awareness about the campaign within their company.

## 5.2. Media-driven triggers of activity

Media and news coverage is essential for any social movement or campaign to gain traction and to maintain its momentum. Media organizations not only provide a larger audience, through the use of digital tools they are able to provide a captivated audience that is willing to participate in online campaigns. Not surprisingly, media organizations played a decisive role in providing initial thrust but also in helping the movement sustain momentum over the observed time frame. The blog post by Anchalee specified that she gave media the permission to use any/all information she had posted. The initial coverage was a large number of organizations but some of them played a larger role. For instance, coverage by Tech Crunch, a favorite of technology workers, was instrumental in bringing about participants to the campaign. Later on, prominent media outlets from abroad including BBC News, The Independent, and The Guardian provided a lot of coverage and participation from the UK.

## 5.3. Industry-driven triggers of activity

Another interesting aspect of the #ILookLikeAnEngineer campaign was the active participation it got from the industry. A range of technology companies – such as Microsoft and Cisco – and others such as GE, Tesla, Bombardier and Union Pacific, participated in the campaign. At one level this is not surprising, the issue of diversity has a disproportionate impact on the engineering and technology workforce and the campaign got started by reactions to billboards placed by a technology company. Still, the role played by companies and their participation without any coordinated effort is noteworthy. Whether to identify or to co-opt the campaign or to show solidarity, companies participated highly. Furthermore, similar to industry's contribution, the campaign also received notable involvement other non-profits and academia (e.g. MIT and CMU) including faculty members, students, and official university accounts and websites.

## 5.4. Personality-driven triggers of activity

The final triggering event type we identified within the dataset was alignment with a known personality. As prior research has shown, there is a positive correlation between users with high number of followers, such as celebrities, on subsequent tweets/retweets [4]. In our data we did see some example of that, for instance, Chelsea Clinton. What we saw even more powerful was the effect of personalities known within this community or related to the effort. For instance, Michelle Sun, a Hong Kong based entrepreneur, who founded First Code Academy that teaches coding to school kids was frequently referred and exemplified with the hashtag #ILookLikeAnEnginner in a number of media stories and Twitter posts. Similarly, a number of news stories, blog posts, and Tweets were supported and amplified the campaign on Ada Lovelace Day celebrated on Oct. 11 each year globally to honor women in Science and Technology and the First female programmer Ada Lovelace.

## 6. Discussion

The analysis of the #ILookLikeAnEngineer Twitter data illustrates that a successful hashtag based campaign for social justice, in this case workplace diversity, was able to leverage a social technology to bring together a diverse range of participants who each played a role in supporting and advancing the cause. Yet, the study illustrates that the campaign did not reside solely in the online world. Although the nature of Twitter made it easy to participate, it took significant effort and in often case planning to keep the movement going. Consequently, there is a broader implication for the role played by social media as a platform to facilitate the engagement between offline and online world. There are always newer issues to attract someone's attention and creating a hashtag has a very low barrier to entry and therefore different strategies are needed to keep people involved and attract new supporters. Our study identifies triggering as one strategic user behavior that supports connective action. The idea of triggering has been used previously in relation to social media to shed light on how to get workers to focus on specific information within organizations so that knowledge sharing is enhanced [38]. In our case the triggering brought back attention to the cause and motivated users – both those who were already involved and new ones – to pay attention to the campaign and participate in it.

The #ILookLikeAnEngineer campaign provided a sense of collective identity to participants, as is the case with most social movement efforts, but it went beyond that in terms of sustaining itself. As our analysis shows, the campaign benefitted from a lot from contribution of different actors, and different kinds of actors, coming together. Not only were there high profile champions of the cause that Tweeted their support, but other personalities that are well known within their communities, and linked to providing access and support to women engineers, were able motivate their supporters to participate. Industry played a pivotal role in the campaign as well. Most engineers end up working in large firms so in some ways this is not surprising. The support these actors provided was not necessarily some form of heavy investment but Tweets and Retweets more in keeping with the notion of connective action.

Also consistent with prior work, social media affordances allowed for participation that was both about the campaign and also relational. Although we do not classify users according to their roles [e.g. 36], in our case – given our larger and more comprehensive dataset, we would probably be looking at a much broader range of roles – we find that either through active collaboration or accidentally, without active planning, different users ended up working in support of each other. These users were not necessarily individuals but also organizations. Consistent with [22] and [23], we found that online amplification of the offline events was facilitated via social media and played a critical role in success of channeling momentum over time. Some of these actions were designed, but not necessarily in terms of when they occurred. The ability to link the hashtag with other efforts, such as Ada Lovelace day, boosted its use. In some ways, it allowed a persistent narrative to emerge not just within Tweets but across the campaign [25]. At this point the hashtag has enough credibility that it continues to be used and associated with STEM education and workforce development related efforts.

Furthermore, the campaign was directed at the technology industry and therefore it was critical for the industry to respond. There was also, though, another reason. The campaign in some ways got co-opted by the industry as a way to improve their public image and in many cases as a recruitment effort. The PR effort is perceivable across a lot of tweets and especially in efforts such as those by Tesla where one of their engineers wrote an article that appeared online in Fortune and which referred to the campaign. Similarly, many of the personalities associated with the campaign provided their support because they believed in the cause – improve gender equity and diversity. Yet, there were also people who leveraged the campaign to better their image by appearing to be on the right side of an issue. Although the co-optation by the industry and even by individuals is problematic, it does not necessarily take anything away from the mission of the movement. There were other participants though whose motivations were purely to leverage the campaign for marketing. In some ways, this is not unique to this campaign but is something that is evident across such campaigns. For instance, Miranda et al. in their study of the SOPA discourse raise a related question of whether social media movements are emancipatory or hegemonic and they find that they are both [34].

The takeaway from the analysis though, in addition to the four triggers we identify, is the uniquely decentralized nature of actors who participated and the activities associated with the movement. Although the movement was being championed by those who started it, there was no centralized coordination and yet after every few weeks or so some action occurred that sustained the momentum of the movement. In some ways this explains in part one of the Vaast et al. [36] findings that "A consideration of connective affordances can help shed light onto the collective use of technology and can help understand why a collective outcome may not be reached even as a technology seems to be heavily used by many potential users (e.g., if an emerging role and its corresponding pattern of feature use is missing or not as active) (p. 21-22)."

Overall, it has contributed to the social movement directed towards increasing the participation of women in engineering and technology. By highlighting the different role women play and the different companies in which they work, and in which they have succeeded, the movement has provided a much diverse image of engineering. It has also provided a model for other such movements and activism efforts, such as #ILookLikeaSurgeon, which is significant from the perspective of how movements inspire and sustain each other on social media platforms. In particular, for those who are looking to start such movements or leverage social media for similar efforts it provides a blueprint to not only start an effort but sustain through its post-initial slump that invariably follows the initial burst.

In spite of its popularity, digital activism, especially hashtag activism has been criticized as a passive activity with very low barrier to entry, relatively effortless participation, and often the absence of any tangible or quantifiable results [11]. This lack of accountability for outcomes is also an issue with #ILookLikeAnEngineer, and suggests a direction for future work. Even though the barrier to participation is slightly higher than just tagging, most users posted a picture holding a sign that referred to the hashtag it is still lower than physically traveling and participating in a march, for instance. Yet, as we illustrated above the continuing popularity of the hashtag and the participation of women in coding events that were started as a result of the campaign, especially through Women Who Code ™ provide some evidence of impact.

## 7. Conclusion

The findings from our study of a Twitter hashtag campaign for gender equity demonstrate how a successful movement goes beyond the initial spike in participation to maintain momentum through its early period. Specifically, we identify four triggers that helped sustain the connective action: 1) Event-Driven: Alignment of the campaign with physical events related to the issue; 2) Media-Driven: News coverage of the event in the media; 3) Industry-Driven: Web participation in the campaign by large organizations; and 4) Personality-Driven: Alignment of the event with popular and/or known personalities. These findings are applicable in other similar efforts as well.

## 8. Acknowledgements

This research was partially supported by U.S. National Science Foundation Awards: 1424444 and 1707837.